\documentclass{emulateapj}
\usepackage{apjfonts}
\lefthead{KIM ET AL.}
\righthead{LIMITS OF MICROLENING BINARIES}

\begin{document}
\title{Limits of Binaries That Can Be Characterized by Gravitational Microlensing}

\author{
Doeon Kim\altaffilmark{1},
Yoon-Hyun Ryu\altaffilmark{2},
Byeong-Gon Park\altaffilmark{3},
Heon-Young Chang\altaffilmark{2},
Kyu-Ha Hwang\altaffilmark{1},\\
Sun-Ju Chung\altaffilmark{3},
Chung-Uk Lee\altaffilmark{3},
and
Cheongho Han\altaffilmark{1,4}\\
}

\altaffiltext{1}{Department of Physics, Institute for Basic Science
Research, Chungbuk National University, Chongju 361-763, Korea}
\altaffiltext{2}{Department of Astronomy and Atmospheric Sciences, 
Kyungpook National University, Daegu 702-701, Korea}
\altaffiltext{3}{Korea Astronomy and Space Science Institute, Hwaam-Dong,
Yuseong-Gu, Daejeon 305-348, Korea}
\altaffiltext{4}{corresponding author}

\submitted{Submitted to The Astrophysical Journal}

\begin{abstract}
Due to the high efficiency of planet detections, current microlensing 
planet searches focus on high-magnification events. High-magnification 
events are sensitive to remote binary companions as well and thus a 
sample of wide-separation binaries are expected to be collected as a 
byproduct.  In this paper, we show that characterizing binaries for 
a portion of this sample will be difficult due to the degeneracy 
of the binary-lensing parameters.  This degeneracy arises because the 
perturbation induced by the binary companion is well approximated by 
the Chang-Refsdal lensing for binaries with separations greater than 
a certain limit.  For binaries composed of equal mass lenses, we find 
that the lens binarity can be noticed up to the separations of $\sim 60$ 
times of the Einstein radius corresponding to the mass of each lens.
Among these binaries, however, we find that the lensing parameters can 
be determined only for a portion of binaries with separations less than 
$\sim 20$ times of the Einstein radius.
\end{abstract}



\section{Introduction}

Searches for extrasolar planets by using microlensing are being 
carried out by observing stars located toward the Galactic bulge 
field.  The lensing signal of a planet is a short-duration perturbation 
to the smooth standard light curve of the primary-induced lensing 
event occurring on a background star \citep{mao91, gould92}.  For 
the detections of the short-duration planetary lensing signals, 
these searches are using a combination of survey and follow-up 
observations, where the survey observations (e.g., OGLE, 
\citet{udalski03}; MOA, \citet{bond02}) aim to maximize the 
lensing event rate by monitoring a large area of sky and the 
follow-up observations (e.g., PLANET, \citet{albrow01}; MicroFUN, 
\citet{dong06}) intensively monitor the events alerted by the survey 
observations.  However, the limited number of telescopes restricts 
the number of events that can be followed at any given time and thus 
priority is given to those events that will maximize the planetary 
detection probability.  Currently, the highest priority is given to 
high-magnification events because the source trajectories of these 
events always pass close to the perturbation region around the 
planet-induced caustic located near the primary lens \citep{griest98}.  
In addition, follow-up observations can be prepared for these events 
because the perturbation typically occurs near the peak of the event, 
which can be predicted from the data on the rising part of the event 
light curve.  As a result, six (OGLE-2005-BLG-071Lb, OGLE-2005-BLG-169Lb, 
OGLE-2006-BLG-109Lb,c, MOA-2007-BLG-400b, MOA-2007-BLG-192) of the 
eight reported microlensing planets were detected through the channel 
of high-magnification events \citep{bond04, udalski05, beaulieu06, 
gould06, gaudi08, dong08, bennett08}.

In addition to planets, high-magnification events are sensitive to 
wide-separation binary companions as well.  Similar to the planetary 
case, the companion of a wide-separation binary induces a small 
caustic close to the primary lens and thus can produce a short-duration 
perturbation near the peak of a high-magnification event.  Due to the 
different nature of the companions, however, the perturbations induced 
by a wide-separation binary companion and a planet can be distinguished
\citep{albrow02, han08}. Therefore, under the current planetary lensing 
strategy focusing on high-magnification events, a considerable number 
of wide-separation binaries are expected to be detected as a byproduct 
and this sample might provide useful information about the physical 
distribution of binaries.

Unlike this expectation, however, we find that characterizing binaries 
for a significant portion of the binary lens sample will be difficult 
due to the degeneracy of the binary-lensing parameters.  This degeneracy 
arises because the perturbation induced by the binary companion is 
well approximated by the Chang-Refsdal lensing (hereafter C-R lensing) 
for binaries with separations greater than a certain value.  The C-R 
lensing represents single-mass lensing superposed on a uniform 
background shear.  For a wide-separation binary, the shear results 
from the combination of the binary-lens parameters and thus the 
individual parameters cannot be separately determined.

The paper is organized as follows.  In \S\ 2, we briefly describe 
the properties of binary and C-R lensing. In \S\ 3, we demonstrate 
the proximity between the binary and C-R lensing for binaries with 
separations beyond a certain limit.  We then set the range in the 
binary-lensing parameter space where the degeneracy of binary-lensing
parameters occurs.  We discuss the meaning of this degeneracy in the 
studies of binaries by using microlensing. We summarize the results 
and conclude in \S\ 4.

\section{Lensing Properties}

If a source is lensed by a single point mass, the lens mapping 
from the lens plane to the source plane is expressed by the lens 
equation
\begin{equation}
\zeta = z - {1\over \bar{z}},
\label{eq1}
\end{equation}
where $\zeta=\xi+i\eta$ and $z=x+iy$ are the locations of the 
source and the lens in complex notations, respectively, and 
$\bar{z}$ denotes the complex conjugates of $z$.  Here all 
angles are normalized by the angular Einstein radius of the 
lens, which is related to the physical parameters of the lens by
\begin{equation}
\theta_{\rm E}=\left({4GM\over c^2}\right)^{1/2}
\left({1\over D_L}-{1\over D_S}\right)^{1/2},
\label{eq2}
\end{equation}
where $M$ is the mass of the lens and $D_L$ and $D_S$ are the 
distances to the lens and source, respectively.  Solving the 
lens equation results in two solutions of the image position.  
The lensing process conserves the source surface brightness and 
thus the magnifications of the individual images correspond to 
the area ratios between the images and source.  Mathematically, 
this is obtained by solving the Jacobian of the mapping equation 
evaluated at the image position, i.e., 
\begin{equation}
A_i=\left\vert 
\left( 
1-{\partial\zeta\over \partial\bar{z}}
{\overline{\partial\zeta}\over \partial\bar{z}}\right)_{z=z_i}^{-1}
\right\vert.
\label{eq3}
\end{equation}
Then, the total magnification corresponds to the sum of the individual 
images, i.e., $A=\sum_i A_i$, and this results in 
\begin{equation}
A={u^2+2\over u(u^2+4)^{1/2}},
\label{eq4}
\end{equation}
where $u=|\zeta|$ represents the normalized lens-source separation.
For a rectilinear motion, the lens-source separation is related to 
the lensing parameters of the Einstein time scale $t_{\rm E}$, the 
closest lens-source separation $u_0$, and the time at the moment $t_0$ 
by 
\begin{equation}
u=\left[ u_0^2+\left( {t-t_0\over t_{\rm E}}\right)^2\right]^{1/2}. 
\label{eq5}
\end{equation}
The light curve of a single-lensing event is characterized by a 
smooth and symmetric shape \citep{paczynski86}.

If a source is lensed by a binary lens, on the other hand, the 
mapping equation is expressed as \citep{witt90}
\begin{equation}
\zeta = z - {m_1/M \over \bar{z}-\bar{z}_{L,1}} 
- {m_2/M \over \bar{z}-\bar{z}_{L,2}},
\label{eq6}
\end{equation}
where $z_{L,1}=x_{L,1}+iy_{L,1}$ and $z_{L,2}=x_{L,2}+iy_{L,2}$ 
are the positions of the lens components, $m_1$ and $m_2$ are 
their masses, and $M=m_1+m_2$ is the total mass of the binary.  
One important characteristic of binary lensing that differentiates 
from single lensing is the formation of caustics, which represent 
source positions at which the magnification of a point source becomes 
infinite. The set of caustics form closed curves, each of which is 
composed of concave curves (fold caustic) that meet at points (cusps).  
Binary lenses can have one, two, or three closed caustic curves. If 
the two masses are separated by approximately an Einstein radius, 
then there is a single six-cusp caustic.  If the masses are much 
closer than an Einstein ring, there is a central four-cusp caustic 
and two outlying three-cusp caustics.  If they are separated by much 
more than an Einstein ring (wide-separation binary), then there are 
two four-cusp caustics, where each of the caustics is associated with 
each member of the binary.  In addition to the parameters of single 
lensing, modelling a binary-lensing event requires two additional 
lensing parameters of the separation $s=|z_{L,1}-z_{L,2}|$ and the 
mass ratio $q=m_2/m_1$ between the lens components.

The Chang-Refsdal lensing represents single lensing superposed 
on a uniform background shear \citep{chang79, chang84}.  The lens 
equation for the C-R lensing is represented by 
\begin{equation}
\zeta=z-{1\over\bar{z}} + \gamma\bar{z},
\label{eq7}
\end{equation}
where $\gamma$ is the shear.  The shear induces a single set of 
caustics, which form around the lens.  The caustic has a shape 
of a hypocycloid with four cusps.  Its size as measured by the 
separation between two confronting cusps are related to the 
shear by
\begin{equation}
\Delta\zeta_c =4\gamma. 
\label{eq8}
\end{equation}

In the limiting case of a binary lens where the binary separation 
is much larger than the Einstein radius ($s\gg 1$), the lensing 
properties in the vicinity of each of the binary lenses is well 
described by that of the C-R lensing \citep{dominik99}.  The shear 
exerted by the companion is related to the binary parameters by
\begin{equation}
\gamma \sim {q\over \hat{s}^2};\qquad \hat{s}=(1+q)^{1/2}s,
\label{eq9}
\end{equation}
where $\hat{s}$ is the binary separation normalized by the Einstein 
radius corresponding to the mass of the lens close to which the 
caustic is located.  The shear decreases as $1/\hat{s}^2$ and 
thus the caustics of a wide-separation binary lens shrink rapidly 
as the binary separation increases.  In the limiting case where 
$\hat{s} \rightarrow \infty$, the shear and the caustics vanish 
and the individual lens components behave as if they are two 
independent single lenses.  The positions of the lens components 
effectively working as single lenses, effective lens position 
$z_{L,eff}$, are slightly shifted from their original positions 
with an offset 
\citep{distefano96}
\begin{equation}
\Delta z_L = z_L - z_{L,eff}
\sim -{q\over \hat{s}} {z_{L,2}-z_{L,1} \over |z_{L,2}-z_{L,1}|}.
\label{eq10}
\end{equation}

\begin{figure*}[t]
\epsscale{1.0}
\plotone{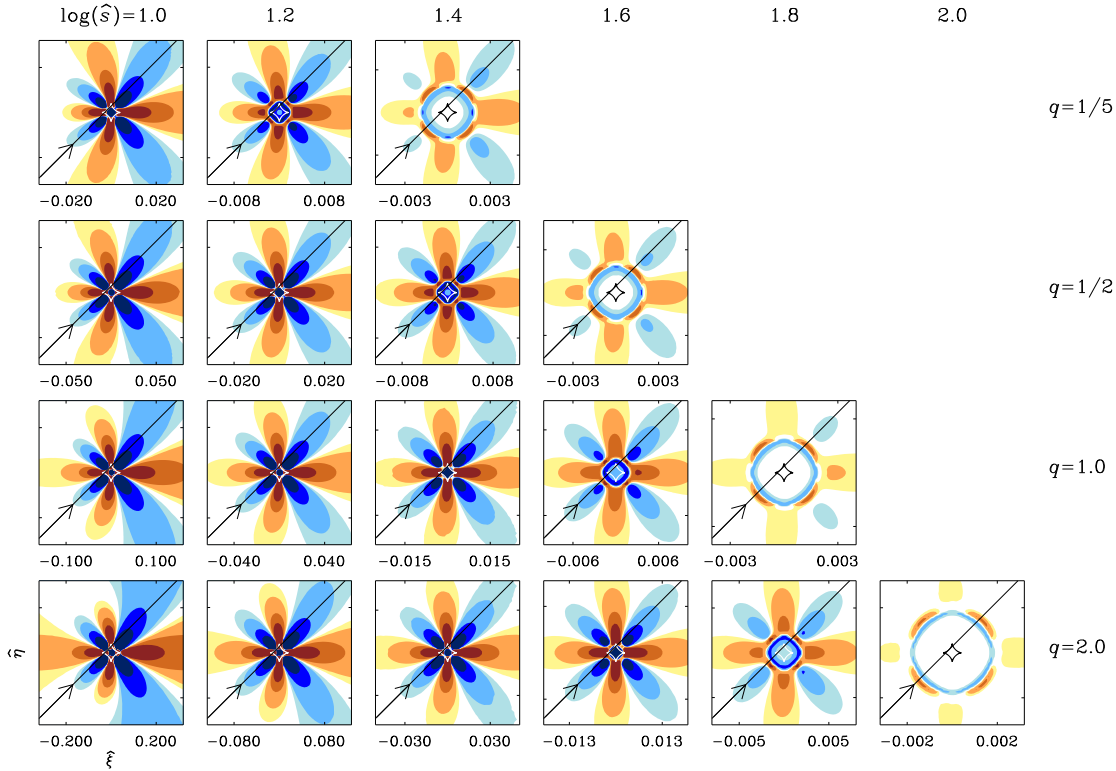}
\caption{\label{fig:one}
Maps of the fractional magnification residual from that of 
single lensing as a function of the source position in the 
central region of a component of binary lenses with various 
separations and mass ratios.  All lengths are scaled by the 
Einstein radius corresponding to the mass of the binary 
component around which the map is constructed (primary).  
The coordinates are centered at the effective position of 
the primary lens (see the definition of the `effective lens 
position' in the text of \S\ 2).  The width of each map 
corresponds to 8 times of the width of the caustic.  In each 
map, the regions with blue and brown-tone colors represent 
the areas where the binary-lensing magnification is lower 
and higher than the single-lensing magnification, respectively.  
For each tone, the color changes into darker scales when the 
residual is $|\epsilon| \geq 1\%$, 2\%, 5\%, and 10\%, 
respectively.  The straight lines with arrows represent the 
source trajectories and the light curves of the resulting 
events are presented in the corresponding panels of 
Fig.~\ref{fig:three}.  We note that maps are not presented 
if the perturbation from single lensing is severely washed 
out by the finite-source effect and thus the magnification 
pattern of a binary-lensing case is difficult to be 
distinguished from that of a single-lensing case.
}\end{figure*}

\begin{figure*}[t]
\epsscale{1.0}
\plotone{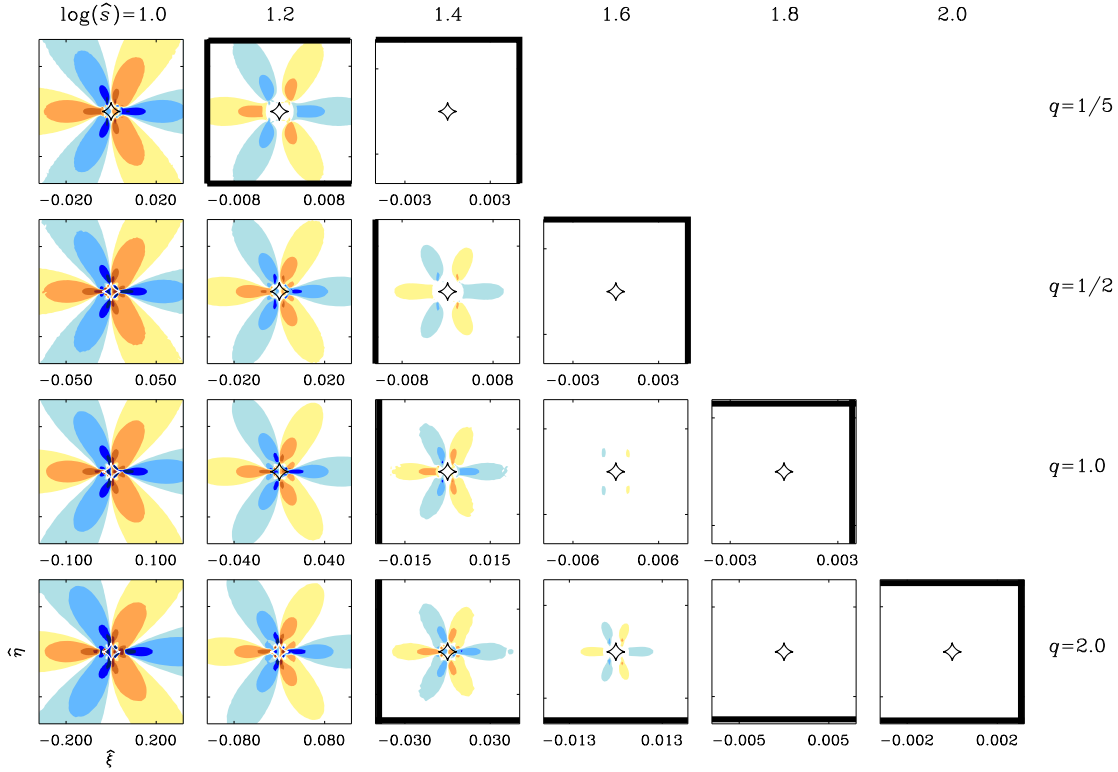}
\caption{\label{fig:two}
Same as Fig.~\ref{fig:one} except that the magnification residual 
is from that of C-R lensing.  The panels blocked by thick lines 
represent the cases where the binary signal can be detected, 
but characterization of the binary is difficult due to the 
degeneracy of the binary-lensing parameters. 
}\end{figure*}

\begin{figure*}[t]
\epsscale{1.0}
\plotone{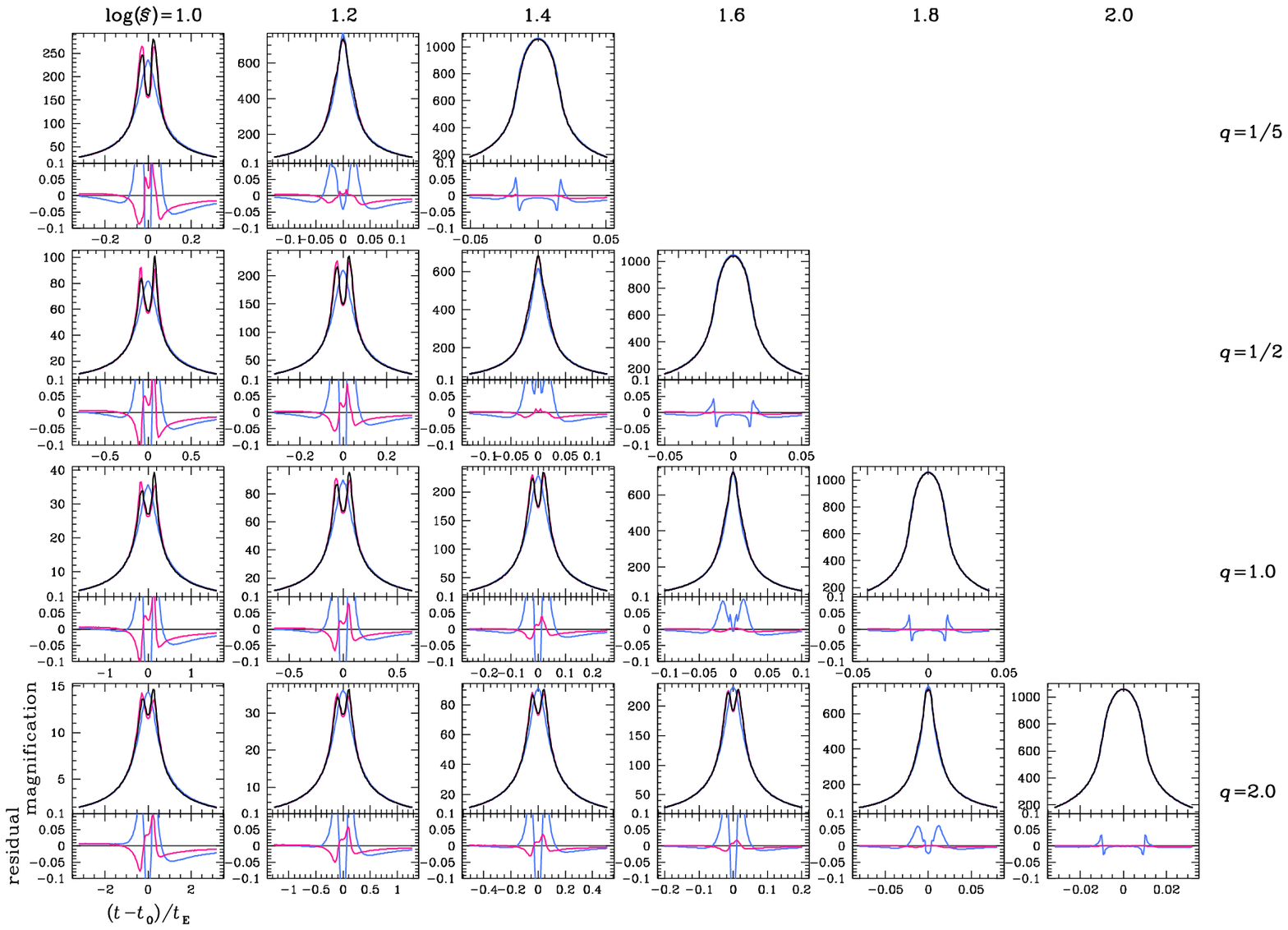}
\caption{\label{fig:three}
Example light curves of high-magnification events caused by 
wide-separation binaries.  The source trajectories responsible 
for the individual events are marked in the corresponding panels 
of Fig.~\ref{fig:one}.  In the upper part of each panel, there 
are three light curves: blue curve for single lensing, red curve 
for C-R lensing, and black curve for binary lensing.  The lower 
part shows the fractional residuals from the single lensing 
magnification (blue curve) and the C-R lensing approximation 
(red curve).
}\end{figure*}

\section{Degeneracy}

The fact that the lensing behavior of a wide-separation binary in 
the region close to each lens component is well described by the 
C-R lensing implies that although lens binarity can be noticed 
from the perturbation to the light curve of a high-magnification 
event, it may be difficult to determine the binary-lensing parameters 
from the analysis of the perturbation.  This is because the C-R 
lensing is described by a single parameter of $\gamma$, which 
results from the combination of the two binary-lensing parameters 
of $s$ and $q$.

Then, naturally rising questions are (1) how serious this degeneracy
is and (2) in which region of the binary-lensing parameter space 
this degeneracy occurs.  To answer these questions, we construct 
two sets of perturbation pattern maps as a function of the source 
position in the central region of a component of binary lenses 
with various separations and mass ratios.  In the first set, we 
construct the maps of the fractional magnification residual from 
that of single lensing, i.e., 
\begin{equation}
\epsilon_s = {A-A_s\over A_s},
\label{eq11}
\end{equation}
where $A$ and $A_s$ are the magnifications of binary and single 
lensing without the companion, respectively.  In the second set, 
we construct the maps of the residual from that of C-R lensing, i.e., 
\begin{equation}
\epsilon_{\rm C-R} = {A-A_{\rm C-R}\over A_{\rm C-R}},
\label{eq12}
\end{equation}
where $A_{\rm C-R}$ is the magnification obtained from C-R 
lensing by approximating the shear exerted by the companion as 
$\gamma=q/\hat{s}^2$.

Figure~\ref{fig:one} and \ref{fig:two} show the two sets of maps. 
In each map, the coordinates are centered at the effective position 
of a binary component.  The width of each map is set such that it 
corresponds to 8 times of the width of the C-R lensing caustic 
computed by equations (\ref{eq8}) and (\ref{eq9}).  The separation 
and the mass ratio of the binary for each map are marked at the 
top and on the right side of the panel, respectively.  The mass 
ratio is such that $q<1.0$ when the companion is less massive 
than the lens component around which the map is constructed and 
vice versa.  In each map, the regions with blue and brown-tone 
colors represent the areas where the binary-lensing magnification 
is lower and higher than the single or C-R lensing magnification, 
respectively.  For each tone, the color changes into darker scales 
when the residual is $|\epsilon| \geq 1\%$, 2\%, 5\%, and 10\%, 
respectively. Figure~\ref{fig:three} shows light curves of events 
resulting from the source trajectories marked in Figure~\ref{fig:one}.  
In the upper part of each panel, there are three curves: blue curve 
for single lensing, red curve for C-R lensing, and black curve for 
binary lensing.  The lower part shows the fractional residuals from 
the single-lensing magnification (blue curve) and the C-R lensing 
approximation (red curve).

The caustic induced by a wide-separation binary is small and 
thus perturbations induced by the caustic are vulnerable to 
finite-source effect.  The finite-source effect is parameterized 
by the ratio of the source radius $r_\star$ to the Einstein 
radius.  For a lensing event toward the Galactic bulge field, 
this ratio is scaled by the physical parameters of the lens by
\begin{equation}
\rho_\star = 9\times 10^{-4} 
\left( {r_\star\over R_\odot}\right)
\left[
\left( {m_1\over 0.3\ M_\odot}\right)
\left( {D_L\over 6\ {\rm kpc}}\right)
\left( 1-{D_L\over D_S}\right)\right]^{-1/2}.
\label{eq13}
\end{equation}
Then, the magnification affected by the finite-source effect
becomes 
\begin{equation}
A={\int_0^{\rho_\star} I(r)A_p(|{\bf r}-{\bf r}_L|)r dr
\over\int_0^{\rho_\star} I(r)r dr},
\label{eq13}
\end{equation}
where ${\bf r}_L$ is the displacement vector of the source center 
with respect to the lens, ${\bf r}$ is the vector to a position 
on the source star surface with respect to the center of the 
source star, $I(r)$ represents the source brightness profile, 
and $A_p$ is the point-source magnification.  We consider the 
finite-source effect by assuming that the source star has a 
uniform disk with a radius equivalent to the Sun and the physical 
parameters of the lens system are $m_1=0.3\ M_\odot$, $D_L=6$ kpc, 
and $D_S=8$ kpc by adopting the values of a typical event being 
detected toward Galactic bulge field.  This results in $\rho_\star 
= 1.8\times 10^{-3}$.  In Figure \ref{fig:one} -- \ref{fig:three}, 
we do not present maps and light curves if the perturbation from 
single lensing is severely washed out by the finite-source 
effect and thus the magnification pattern of a binary-lensing 
case is difficult to be distinguished from that of a single-lensing 
case.

The residual map from single lensing shows that although the 
perturbation becomes weaker as the distance to the companion 
increases, it lasts for a considerable distance.  The maximum 
distance for the detection of the companion signal depends on 
the companion/primary mass ratio and the observational precision.  
By adopting a threshold residual of $\epsilon_s=5\%$, we find 
that companions can be detected up to the separations of 
$\hat{s}\sim 31$, 50, 63, and 100 for binary companions with 
mass ratios of $q=1/5$, 1/2, 1.0, and 2.0, respectively.  
Considering that the physical radius of the Einstein ring is 
$r_{\rm E}\sim 1.9$, AU, these limits correspond to the physical 
separations of $d\sim 59$, 95, 120, and 190 AU, respectively, 
implying that high-magnification events provide an efficient 
channel to detect wide-separation binaries.

However, the condition for detection is different from the 
condition for characterization.  The condition for binary 
characterization depends on various factors such as real 
experimental data, photometric errors, sampling rate, 
observation duration, and telescopes performance.  Therefore, 
it is not easy to define a single observable that can be 
used for an indicator to judge the characterizability of 
binary-lens events.  However, if a light curve significantly 
deviates from C-R lensing approximation, it would be 
possible to determine the binary-lensing parameters.  The 
photometric error achieved by the current follow-up observations 
reaches down to $\sim 1-2\%$ level for high-magnification events 
(Gould 2008, private communication). We therefore set the 
condition for binary characterization such that light curves
should be distinguished from the C-R lensing approximation with 
$\epsilon_{\rm C-R}\geq 5\%$.  With this criterion, we find 
that the upper limits of binary separation for characterization 
are $\hat{s}\sim 13$, 20, 22, and 25 for binary companions with 
mass ratios of $q=1/5$, 1/2, 1.0, and 2.0, respectively.  From 
the comparison of the limits for detection, the ranges of binary 
separation for characterization are substantially narrower than 
the ranges for detection. This implies that for a significant 
fraction of events for which signals of wide-separation binary 
companions are detected, it will be difficult to characterize 
the properties of the binary by determining the lensing parameters.  
In Figure~\ref{fig:two}, we mark this region of degeneracy by 
blocking the panels with thick solid lines.  We also list the 
detection and degeneracy regions in Table~\ref{table:one}.

\begin{deluxetable}{lll}
\tablecaption{Detection and Degeneracy Zones\label{table:one}}
\tablewidth{0pt}
\tablehead{
\multicolumn{1}{c}{mass } &
\multicolumn{1}{c}{detection} &
\multicolumn{1}{c}{degeneracy}\\
\multicolumn{1}{c}{ratio} &
\multicolumn{1}{c}{zone} &
\multicolumn{1}{c}{zone}
}
\startdata
$q=1/10$ &  $s \lesssim 24$  ($d \lesssim 46\ {\rm AU}$)  &  $11 \lesssim s \lesssim 24$  ($21\ \lesssim d \lesssim 46\ {\rm AU}$)  \\
$q=1/5$  &  $s \lesssim 31$  ($d \lesssim 59\ {\rm AU}$)  &  $13 \lesssim s \lesssim 31$  ($25\ \lesssim d \lesssim 59\ {\rm AU}$)  \\
$q=1/3$  &  $s \lesssim 40$  ($d \lesssim 76\ {\rm AU}$)  &  $16 \lesssim s \lesssim 40$  ($20\ \lesssim d \lesssim 76\ {\rm AU}$)  \\
$q=1/2$  &  $s \lesssim 50$  ($d \lesssim 95\ {\rm AU}$)  &  $20 \lesssim s \lesssim 50$  ($38\ \lesssim d \lesssim 95\ {\rm AU}$)  \\
$q=1.0$  &  $s \lesssim 63$  ($d \lesssim 120\ {\rm AU}$) &  $22 \lesssim s \lesssim 63$  ($41\ \lesssim d \lesssim 120\ {\rm AU}$) \\ 
$q=2.0$  &  $s \lesssim 100$ ($d \lesssim 190\ {\rm AU}$) &  $25 \lesssim s \lesssim 100$ ($48\ \lesssim d \lesssim 190\ {\rm AU}$)
\enddata
\tablecomments{
Detection and degeneracy zones of wide-separation binaries.
The detection zone represents the range of binary separation
where the lensing signature of a companion can be detected.
The degeneracy zone indicates the range where the companion 
can be detected but the characterization of the binary is 
difficult due to the degeneracy of the binary-lensing parameters. 
The absolute physical ranges are given assuming a microlensing 
system with $D_S=8\ {\rm kpc}$, $D_L=6\ {\rm kpc}$, and $m_i=
0.3\ M_\odot$.
}
\end{deluxetable}

\section{Conclusion}

By investigating the patterns of central perturbations produced 
by wide-separation binary companions, we found that the perturbation 
can be noticed and thus lens binarity can be identified for 
binaries with considerable separations.  However, we also found 
that perturbations for a significant portion of these binaries 
can be well approximated by the C-R lensing, implying that the 
individual binary-lensing parameters cannot be separately determined 
and thus characterization of the binary is difficult.  We set the 
range in the binary-lensing parameter space where this degeneracy 
occurs.  From this, we found that the lens binarity can be noticed 
up to the separations of $\sim 60$ times of the Einstein radius 
corresponding to the mass of each lens of a binary composed of 
equal mass lenses.  Among these binaries, however, we found that 
the lensing parameters can be determined only for a portion of 
binaries with separations less than $\sim 20$ times of the Einstein 
radius.

\acknowledgments 
C.\ H. acknowledge support from the Astrophysical Research Center 
for the Structure and Evolution of the Cosmos (ARCSEC) of the 
Korea Science and Engineering Foundation (KOSEF) through the 
Science Research Program (SRC).  H.-Y.\ C. and B-G.\ P. were 
supported by the grant (KRF-2006-311-C00072) from Korea Research 
Foundation.

\end{document}